\documentclass[prd,letterpaper,superscriptaddress,reprint,nofootinbib]{elsarticle} 
\usepackage{graphicx}
\include{rlFig}
\usepackage{amssymb}
\usepackage{amsthm} 
\usepackage{amsmath}
\usepackage{amsbsy}
\usepackage[caption=false]{subfig}
\usepackage{natbib}
\graphicspath{{./}{fig/}}
\title{Measurement of the real dielectric permittivity $\epsilon_r$ of glacial ice} 

\author[2,3]{Allison, P.}
\author[4]{Archambault, S.}
\author[5]{Auffenberg, J.}
\author[6]{Bard, R.}
\author[2,3]{Beatty, J.J.}
\author[5]{Beheler-Amass, M.}
\author[1,7]{Besson, D.Z.}
\author[5]{Beydler, M.}
\author[8]{Chen, C.C.}
\author[8]{Chen, C.H.}
\author[8]{Chen, P.}
\author[5]{Christenson, A.}
\author[2,3]{Clark, B.A.}
\author[2,3]{Connolly, A.}
\author[9]{Cremonesi, L.}
\author[10]{Deaconu, C.}
\author[5]{Duvernois, M.}
\author[6]{Friedman, L.}
\author[4]{Gaior, R.}
\author[2,3]{Hanson, J.}
\author[5]{Hanson, K.}
\author[5]{Haugen, J.}
\author[6]{Hoffman, K.D.}
\author[2,3]{Hong, E.}
\author[8]{Hsu, S.Y.}
\author[8]{Hu, L.}
\author[8]{Huang, J.J.}
\author[8]{Huang, M.-H. A.}
\author[4]{Ishihara, A.}
\author[5]{Karle, A.}
\author[5]{Kelley, J.L.}
\author[5]{Khandelwal, R.}
\author[4]{Kim, M.-C.}
\author[11]{Kravchenko, I.}
\author[11]{Kruse, J.}
\author[4]{Kurusu, K.}
\author[4]{Kuwabara, T.}
\author[12]{Landsman, H.}
\author[1]{Latif, U.A.}
\author[5]{Laundrie, A.}
\author[8]{Li, C.-J.}
\author[8]{Liu, T.-C.}
\author[5]{Lu, M.-Y.}
\author[4]{Mase, K.}
\author[5]{Meures, T.}
\author[4]{Nam, J.}
\author[9]{Nichols, R.J.}
\author[12]{Nir, G.}
\author[1,7]{Novikov, A.}
\author[10]{Oberla, E.}
\author[5]{O' Murchadha, A.}
\author[13]{Pan, Y.}
\author[2,3]{Pfendner, C.}
\author[1]{Ratzlaff, K.}
\author[3]{Relich, M.}
\author[13]{Roth, J.}
\author[5]{Sandstrom, P.}
\author[13]{Seckel, D.}
\author[8]{Shiao, Y.S.}
\author[11]{Shultz, A.}
\author[6]{Song, M.}
\author[6]{Touart, J.}
\author[14]{Varner, G.S.}
\author[10]{Vieregg, A.}
\author[8]{Wang, M.Z.}
\author[8]{Wang, S.H.}
\author[15]{Wissel, S.}
\author[4]{Yoshida, S.}
\author[1]{Young, R.}
\address[1]{Dept. of Physics and Astronomy, Univ. of Kansas, Lawrence, KS, USA}
\address[2]{Dept. of Physics, The Ohio State University, 191 West Woodruff Avenue, Columbus, OH, USA}
\address[3]{Center for Cosmology and Astro-Particle Physics, The Ohio State University, 191 West Woodruff Avenue, Columbus, OH, USA}
\address[4]{Dept. of Physics, Chiba University, Tokyo, Japan}
\address[5]{Dept. of Physics and Wisconsin IceCube Particle Astrophysics Center, University of Wisconsin, Madison, WI, USA}
\address[6]{Dept. of Physics, Univ. of Maryland, College Park, MD, USA}
\address[7]{National Research Nuclear University, Moscow Engineering Physics Institute, Moscow, Russia}
\address[8]{Dept. of Physics, Grad. Inst. of Astrophys., \& Leung Center for Cosmology and Particle Astrophysics, National Taiwan Univ., Taipei, Taiwan}
\address[9]{Dept. of Physics and Astronomy, Univ. College London, London, United Kingdom}
\address[10]{Dept. of Physics, University of Chicago, Chicago, IL, USA}
\address[11]{Dept. of Physics and Astronomy, Univ. of Nebraska-Lincoln, NE, USA}
\address[12]{Weizmann Institute of Science, Rehovot, Israel}
\address[13]{Dept. of Physics and Astronomy, Univ. of Delaware, Newark, DE, USA}
\address[14]{Dept. of Physics and Astronomy, Univ. of Hawaii, Manoa, HI, USA}
\address[15]{Dept. of Physics, California Polytechnic State University, San Luis Obispo, CA, USA}

\begin{document}

\begin{abstract}
\noindent 
Owing to their small interaction cross-section, neutrinos are unparalleled astronomical tracers. Ultra-high energy (UHE; $E>$ 10 PeV) neutrinos probe the most distant, most explosive sources in the Universe, often obscured to optical telescopes. Radio-frequency (RF) detection of Askaryan radiation in cold polar ice is currently regarded as the best experimental measurement technique for UHE neutrinos, provided the RF properties of the ice target can be well-understood. To that end, the Askaryan Radio Array (ARA) experiment at the South Pole has used long-baseline RF propagation to extract information on the index-of-refraction ($n=\sqrt{\epsilon_r}$) in South Polar ice. Owing to the increasing ice density over the upper 150--200 meters, rays are measured along two, nearly parallel paths, one of which refracts through an inflection point, with differences in both arrival time and arrival angle that can be used to constrain the neutrino properties. We also observe (first) indications for RF ice birefringence for signals propagating along predominantly horizontal trajectories, corresponding to an asymmetry of order 0.1\% between the ordinary and extra-ordinary birefringent axes, numerically compatible with previous measurements of birefringent asymmetries for vertically-propagating radio-frequency signals at South Pole. Qualitatively, these effects offer the possibility of redundantly measuring the range from receiver to a neutrino interaction in Antarctic ice, if receiver antennas are deployed at shallow (z$\sim$-25-- -100 m) depths. Such range information is essential in determining both the neutrino energy, as well as the incident neutrino direction.
\end{abstract}

\maketitle

\section{Introduction}
The glacial ice in Antarctica offers a unique opportunity for detection of neutrinos.
There are currently three Antarctic experiments which seek detection, via the Askaryan effect\citep{Askaryan1962a,Askaryan1962b,Askaryan1965}, of UHE neutrinos using the ice sheet as a neutrino target\citep{halzen2017high}. Hadronic and electromagnetic showers resulting from neutrino collisions with ice molecules acquire, as they evolve, a net negative charge as atomic electrons are Compton scattered into the forward-moving shower and shower positrons depleted via annihilation with atomic electrons, resulting in a coherent, detectable electromagnetic signal at radio wavelength scales, distributed on a Cherenkov cone approximately 1--2 degrees in transverse width, and with half-opening angle $\sim 57^o$.
Above 10 EeV, the most promising (``cosmogenic'') neutrino source for these experiments\citep{engel2001neutrinos,kotera2010cosmogenic,ahlers2012minimal,Asano:2016scy} 
results from photoproduction of pions
due to interactions
of ultra-high energy nucleons with cosmic microwave background (CMB) photons, with subsequent 
decays to neutrinos. Recent measurements by the IceCube experiment of PeV-scale neutrinos\citep{halzen2017high} have intensified interest in a large-scale radio array capable of measuring the continuation of that spectrum at higher energies; the observation of the first identified extra-galactic neutrino source has provided additional impetus\citep{icecube2018multimessenger}. 

\begin{figure}[htpb]
\centerline{\includegraphics[width=0.55\textwidth,angle=0]{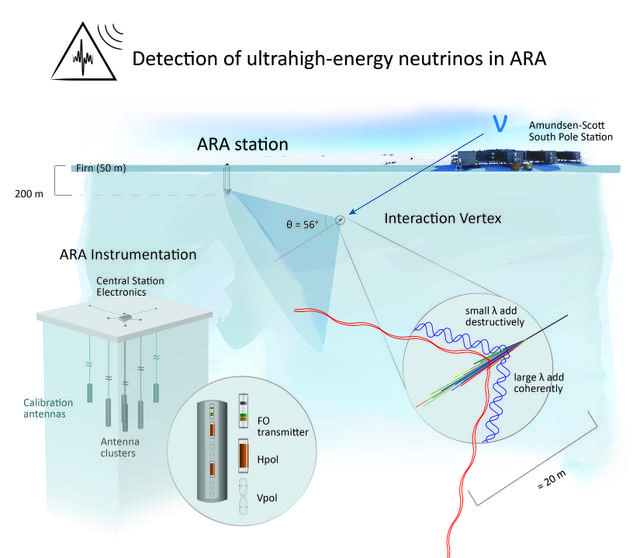}} 
\caption{Schematic of ARA neutrino detection, showing the ARA receiver antennas illuminated by radio-frequency 
Cherenkov signal resulting from an in-ice neutrino interaction.}\label{fig:ara}
\end{figure}

\begin{figure}[htpb]\centerline{\includegraphics[width=0.55\textwidth]{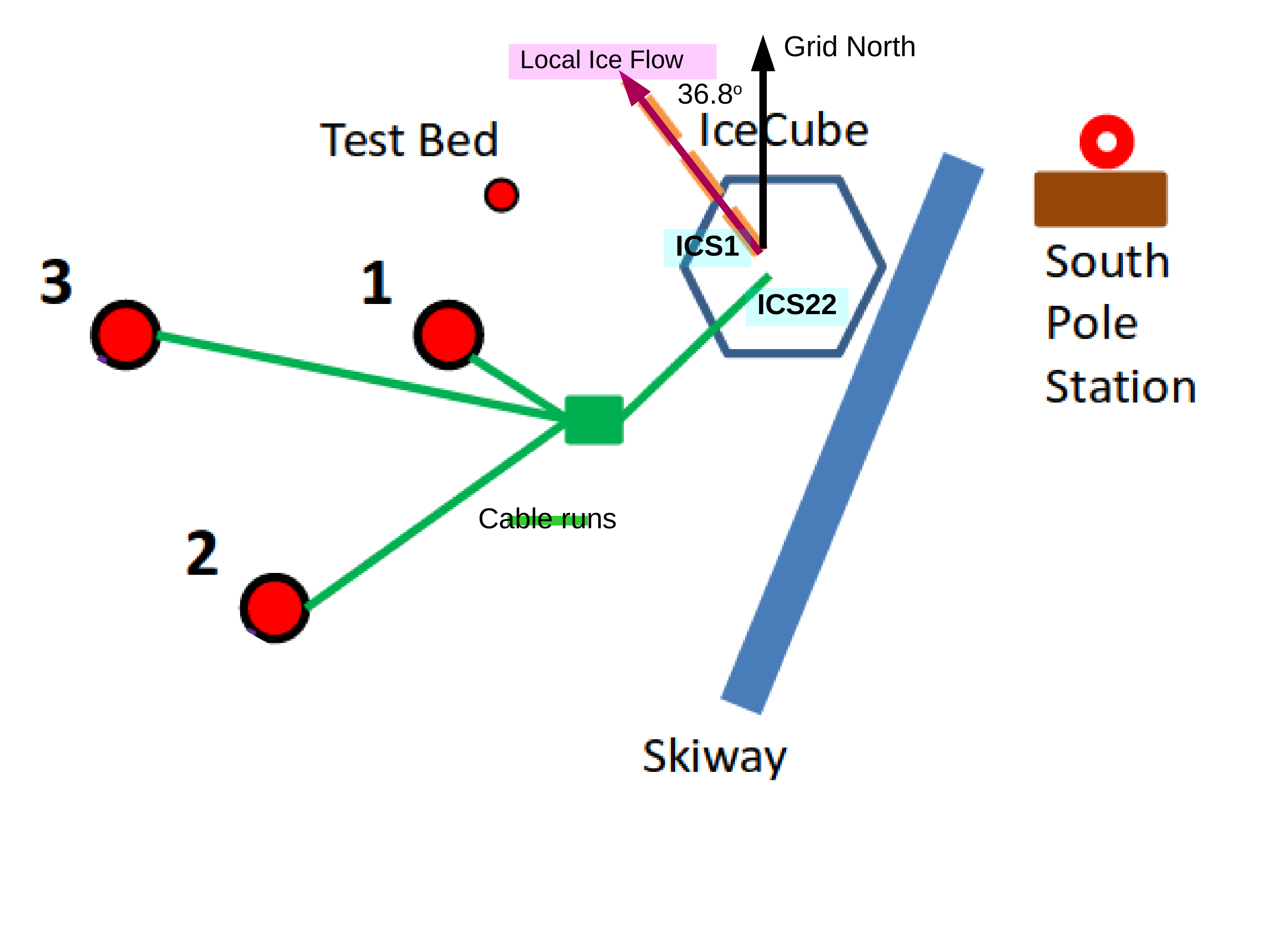}}
\caption{Station geometry for measurements described herein.}\label{fig:a0-a3}
\end{figure}

The Askaryan Radio Array\citep{allison2012design,allison2015first,allison2016performance:2015eky} (ARA) at the South Pole (Figure \ref{fig:ara}) has proposed multiple, independent stations
in a hexagonal
array with inter-station spacing of 2 km. 
Following the initial 2010/11 deployment of a ``TestBed'' in the upper 30 meters of the
South Polar ice sheet, three more (ARA-1, ARA-2, and ARA-3) stations were deployed in 2011/12 and 2012/13, 
with two more (ARA-4 and ARA-5) commissioned in 2017/18, including an advanced trigger system exploiting phased 
array techniques\citep{vieregg2016technique} on ARA-5. 
The analysis described below is based on the pre-2017 station configuration, as shown in Figure \ref{fig:a0-a3}.

\begin{figure}[htpb]\centerline{\includegraphics[width=0.55\textwidth]{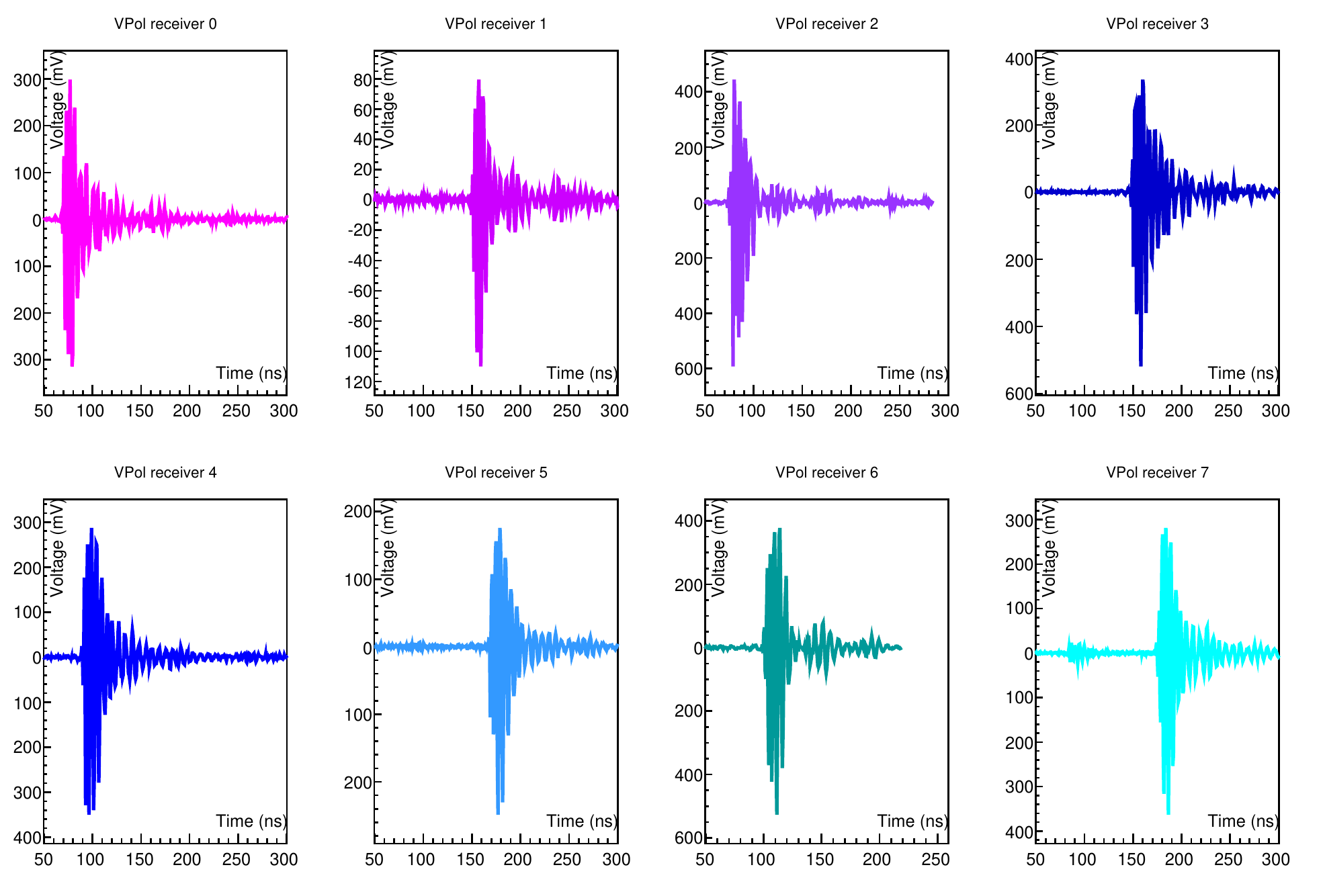}}
\caption{ARA-2 VPol response to the local, VPol calibration pulser, shown for the eight vertically-polarized (VPol) channels. 
HPol signals (not shown) show only noise for these events.}\label{fig:CPev}
\end{figure}

Following ARA-1, for which a drill malfunction
limited deployment to approximately 40\% of the desired 200 meter depth, the subsequent station receivers were deployed at depths of 175-200 m. Each station 
includes 16 antennas, 8 sensitive to vertically polarized (Vpol) and 8 sensitive to horizontally
polarized (Hpol) radiation, with in-ice bandwidths of approximately 150-700 MHz and 250-400 MHz, respectively.
Surface antennas, sensitive over the 25--800 MHz regime, deployed on the TestBed, ARA-1, ARA-2 and ARA-3,
can be used to monitor low-frequency galactic noise, although they rarely give signals coincident with the in-ice antennas,
given the typical time delays in signal
arrival times between the surface and the deep station antennas. In-ice antennas are
installed on four strings defining the corners of a cuboid approximately 20m in height and 20m along the 
horizontal diagonal. An H/V pair is located at each corner of the cuboid, consisting of an Hpol antenna deployed
2m--3m above a Vpol antenna, such that H/V signal arrival times should be synchronous to within 10 ns.
Signals are amplified at the antenna, passed by an in-ice RF-over-optical fiber link to the
surface, and then converted back to RF voltage signals by a surface optical fiber receiver
before entering the data acquisition system.
Full station data collection is triggered whenever (3 of 8 Hpol) .OR. (3 of 8 Vpol) antennas
exceed some voltage threshold within a time window (170 ns) inclusive to the RF travel time across the station.
Thresholds are dynamically adjusted to maintain a
combined event trigger rate of 5-7 Hz, comfortably below the saturation data-taking rate of 25 Hz. 
Following the issue of a valid trigger, signals from all 16 antennas are digitized and
stored, with a readout window typically wide enough to include $\sim$100 ns of pre-trigger and
$\sim$300 ns of post-trigger waveform data. A local calibration pulser, approximately 30 meters displaced from the centroid of the receiver array, 
emits signals at one pulse per second, to monitor
receiver performance. For the testbed, received cal pulser signals in both HPol and VPol are presented elsewhere\citep{allison2012design}. 
The response of the vertically-polarized receivers to the local, vertically-polarized cal pulser, for ARA-2 is shown in 
Figure \ref{fig:CPev}.

ARIANNA\citep{Barwick:2014rca}, located on the Ross Ice Shelf, features an isolated, radio-quiet site with log-periodic dipole antennas deployed on the surface. Downward-pointing antennas search for upcoming Askaryan signal generated by neutrino interactions in the ice; upward-pointing antennas have been used to measure down-coming Askaryan-like `geomagnetic' signals generated by charged cosmic rays interacting in the Earth's atmosphere\citep{barwick2017radio}. The ANITA experiment\citep{GorhamAllisonBarwick2009} features a suite of radio-frequency horn antennas suspended from a balloon flying at an elevation of 38 km in a circumpolar orbit over the Antarctic continent, scanning for upcoming radio signals resulting from charged cosmic ray or neutrino interactions. 

\section{Geometric Optics and Ray tracing}
\begin{figure}[htpb]
\centerline{\includegraphics[width=0.5\textwidth]{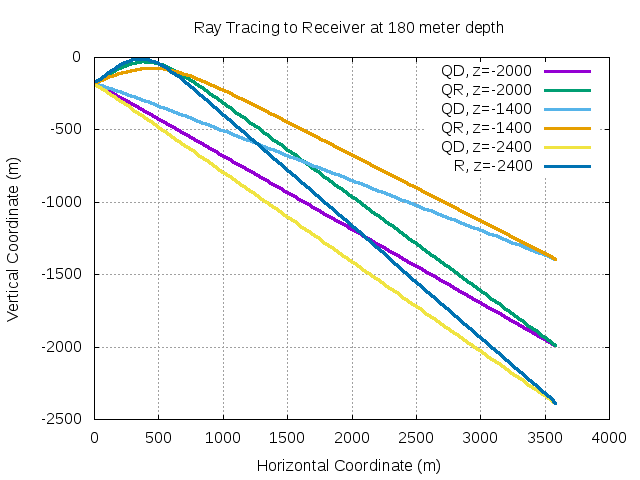}} 
\caption{Simulation of ray tracing assuming ARA index-of-refraction profile with depth (n(z)=1.78-0.43exp(13.2z)), showing trajectories of Quasi-Direct (QD), Quasi-Refracted (QR) rays and surface-reflected (R) rays as a function of lateral distance (x, in meters) and vertical distance (z, in meters). Signals are modeled from three possible source depths (1400m, 2000m, or 2400m) 3600m laterally displaced from a receiver 180m deep. Refracted and direct rays are typically separated by approximately 40 degrees at the measurement point with corresponding launch angles at the source separated by $\sim$5 degrees. By comparison, (QD,QR) launch angle separations for neutrino interactions are limited to two degrees. Note shadow zones in upper right of plot.}\label{fig:RayTracing}
\end{figure}

The sensitivity of any neutrino-search experiment such as ARA
depends on a) the degree of signal absorption in the
target ice medium (determined by the imaginary component of the ice dielectric permittivity), and b) the
volume of ice `visible' to the radio receiver array (determined by the real component of the ice dielectric permittivity).
The absorption length for RF signals in the frequency range of interest (100--1000 MHz) has been measured to exceed 
one km in the upper 1.5 km of the South Polar ice sheet, making it an ideal medium for neutrino
detection\citep{barwick2005south}. However, the changing density of the ice results
in a group velocity varying monotonically with depth. In such a case, Fermat's Least-time principle implies that a) rays will follow curved paths, and b) there may be regions which are `shadowed', for which the superposition of all contributing rays gives zero net amplitude.
Moreover, there may be multiple signals observed from a single source, resulting from either continuous refraction
through the ice itself, or reflection from the upper ice/air interface as illustrated in Figure \ref{fig:RayTracing}. 

The index-of-refraction should itself roughly scale with
the local ice density. Robin suggested the parameterization n(z)=1+0.86$\rho(z)$\citep{robin1975velocity} based on a clever interferometric technique, in which signals recorded on the
surface from a transmitter lowered into a Devon Island ice bore hole were mixed with a fixed frequency; the
wavelength of the ice at the depth of the transmitter was then inferred from the measured beats. 
Similar parameterizations can be found elsewhere (n(z)=0.992+0.848$\rho(z)$, e.g., as determined from a fit to McMurdo Sound ice measurements\citep{kovacs1993reassessment}), although not all obey the expectation that n(z)$\to$1 in the limit of $\rho$=0.

In a recent companion publication, also addressing the question of the index-of-refraction profile at the South Pole and elsewhere on the
Antarctic continent, the general refractive index form is derived from Fermat's Least Time principle\citep{barwick2018observation}. Also presented in that reference are fits to the density data and the inferred best-fit parameters. 
We note that the density data at South Pole show 
considerable variations from experiment-to-experiment, as well as considerable 
deviations from smoothness, which could, in principle, result in sub-dominant `channeling' effects.
This possibility is especially interesting in the context of reports of horizontal
propagation of in-ice RF signals emanating from within the `shadow zone' expected 
in the presence of a gradient to the index of refraction\citep{ShiHaoWang2017,JordanHansonUCIphdthesis,barwick2018observation,deaconu2018measurements}. Such
propagation could occur if there are density layers in the firn. A density inversion could produce
a horizontal waveguide where radiation is confined by refraction, similar to an optical fiber with a
graded index of refraction. 
Similarly, weak discontinuities in density can result in scattering 
surfaces for highly inclined rays, producing a ``channel'' for horizontal propagation. 


\subsection{Deep pulser broadcasts to the ARA array}
In the middle of the expected cosmogenic neutrino energy spectrum
($E_\nu\sim 10^{18.5}$ eV), ARA is designed to detect sources several km distant. It is therefore critical to 
understand the properties of the ice within the array, especially at the 1-2 km depths corresponding to the regime 
over which the bulk of detectable neutrinos are expected to interact. Anticipating
this, during the last year (2010-11) of IceCube construction, two pulsers were deployed on IceCube string
1 at depths z=-1400 m and z=-2450 m (``ICS1'' and ``ICD1'', respectively), 
and one on IceCube string 22 (``ICS22'') at a depth z=-1400 m, proximal to the planned ARA. The pulser at 2450 m depth
was operated in conjunction with the ARA TestBed to validate the South Pole index of refraction 
profile with depth n(z) and also the RF attenuation dependence on depth\citep{allison2012design}, but failed within
the first year of operation. 
The pulsers at depths of 1400 m were operated in
2014/15 as part of the calibration of ARA-2 and ARA-3, and again in 2016/17 after the ARA-2
trigger timing and readout window was adjusted, enabling capture of an extended waveform.
\message{The ARA-3 array was inoperable at the time of the 2017 data-taking.}

Given the monotonic increase of n(z) with depth over
the upper 150 meters of the ice sheet,
the presence of two rays from source to receiver is generic for our geometry (Fig. \ref{fig:RayTracing}), with 
a quasi-direct (QD) ray typically upcoming at the station and either a reflected or a quasi-reflected ray (``QR'', cresting prior to ``true'' reflection at the upper surface) downgoing. 
The maximum in-ice height of the QR ray can be determined from Snell's law and either the
launch angle of the ray at the source, or the received angle at the station. The time delay between arrival of the QD and the QR rays
is of order hundreds of ns; with the exception of the 2016/17 ARA-2 data, this time lag is generally
larger than the waveform capture window. 


\subsection{Analysis of QD and QR rays}

\begin{figure}[htpb]\centerline{\includegraphics[width=0.65\textwidth]{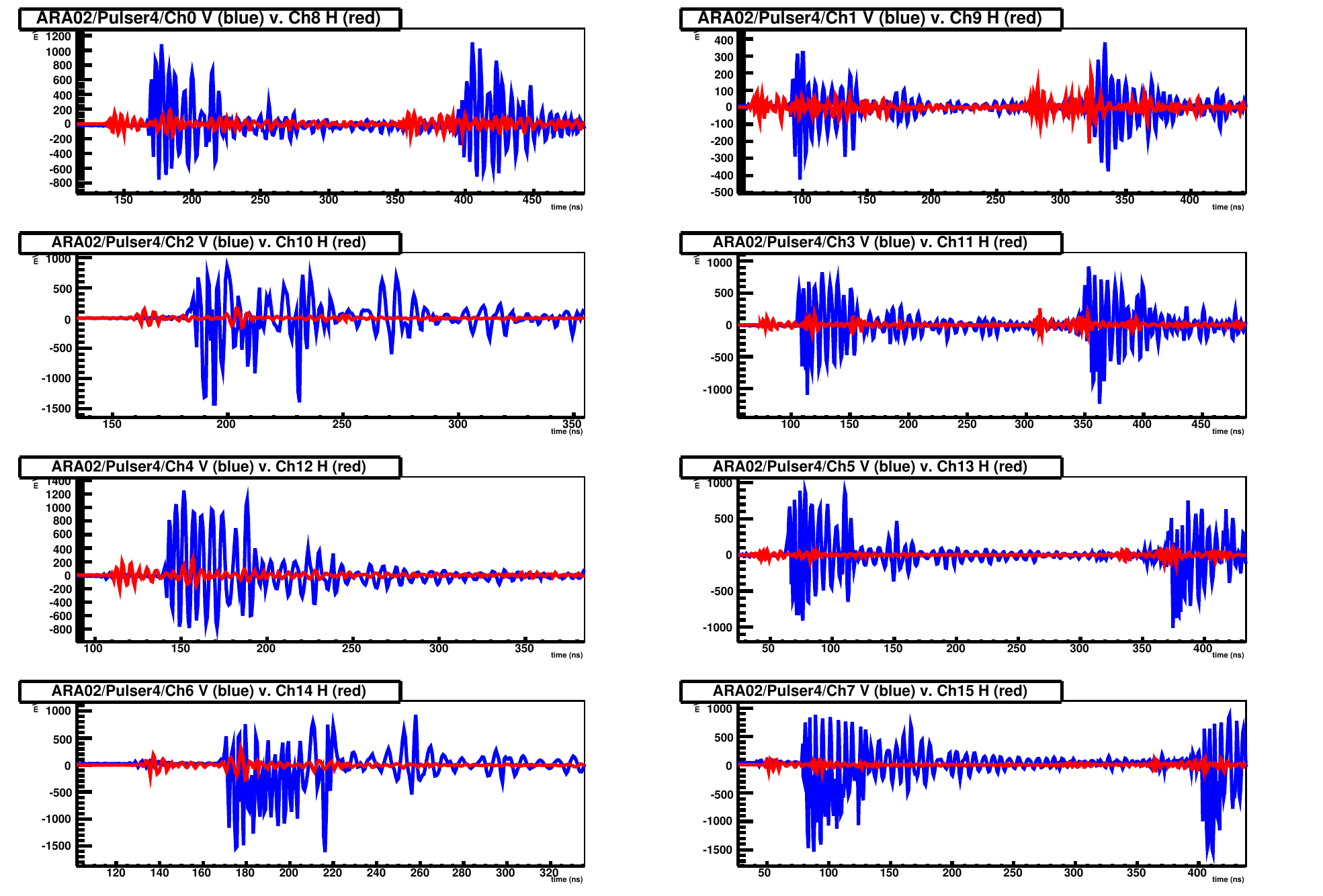}}\caption{ICS1 waveform captures registered by ARA-2 station showing 16 channels in 8 H/V co-located pairs (blue=VPol [Channels 0--7]; red=HPol [Channels 8--15], with significantly smaller amplitude). The QD and QR signals, separated by hundreds of ns, are clearly evident in five of the eight panels. We als observe a time advance of the lower-amplitude HPol channels (red) relative to the VPol channels.}\label{fig:ARA2_DP4}\end{figure} 
\begin{figure}[htpb]\centerline{\includegraphics[width=0.65\textwidth]{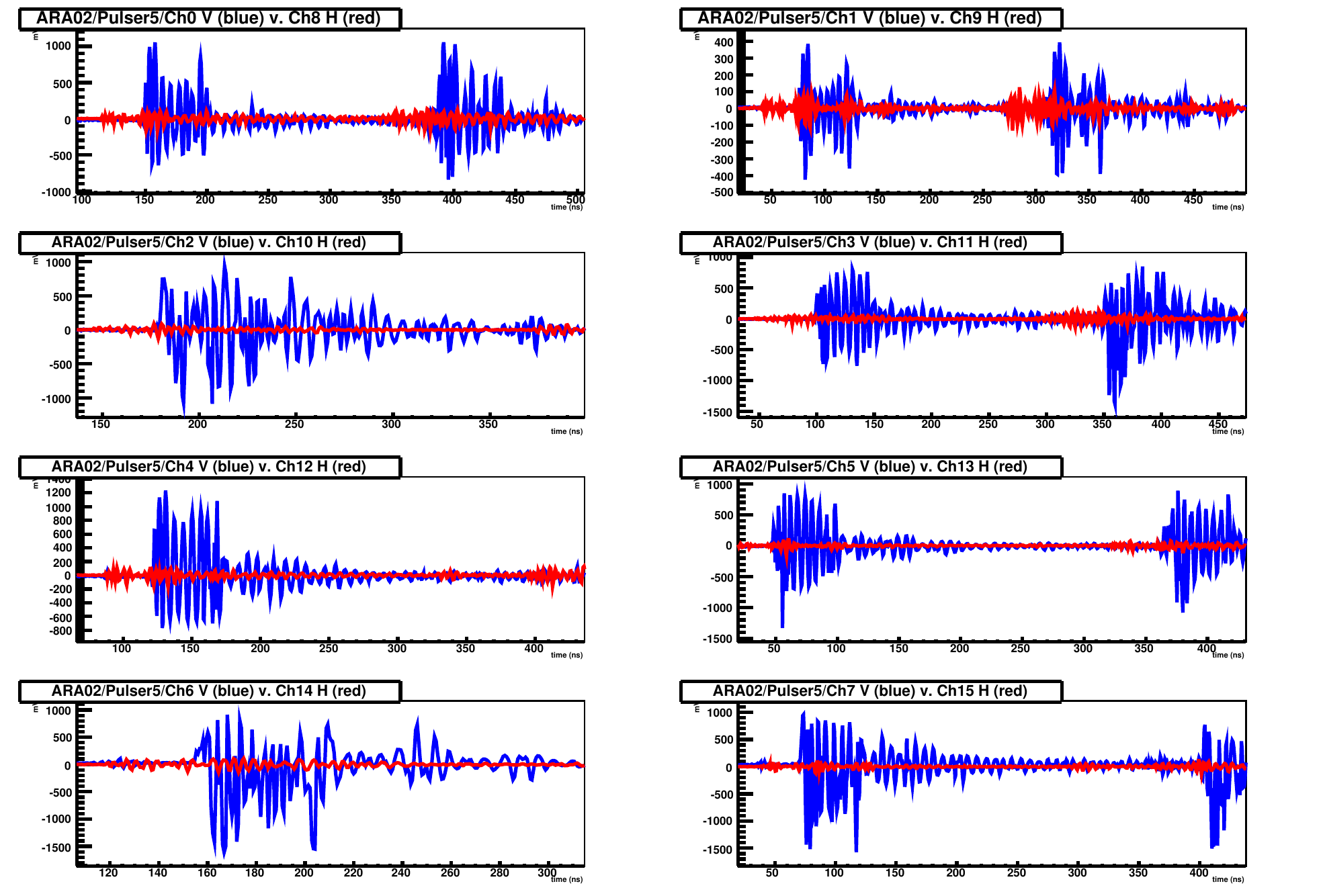}}\caption{ICS22 waveform captures registered by ARA-2 station showing 16 channels in 8 H/V pairs (blue=VPol [Channels 0--7]; red=HPol [Channels 8--15]). }\label{fig:ARA3_DP5}\end{figure} 

Figures \ref{fig:ARA2_DP4} and \ref{fig:ARA3_DP5} show deep pulser events which trigger ARA-2 and ARA-3, respectively. 
The QD rays travel exclusively through deep ice; the sharpness of their leading edge implies relatively little scattering and/or
dispersive effects, consistent with studies of vertical echoes\citep{Besson:2010ww}. 
By contrast, the pulses identified as QR rays suggest a precursor which, we speculate, may be due to scattering
in the firn layer near the top of the pulse trajectory. 
The somewhat irregular signal shapes from the deep pulser have been considered in a previous article\citep{allison2012design}.
In addition to evident saturation of the front-end amplifiers, resulting in compression and signal dilation,
we speculate that the transmitters may have suffered damage after deployment\footnote{Unfortunately, the cross-polarization characteristics of the transmitter were not measured prior to deployment.}, resulting in imperfect coupling of the transmitter itself to the 
antenna used to broadcast the signal. Such imperfect coupling can result in loss of fidelity, as well as enhanced
cross-polarization gain.

The observed received HPol signals, having magnitude $\sim$5--10\% relative to VPol, 
from the nominally deep VPol transmitters is somewhat unexpected.
In addition to possible HPol emission at the source,
such an effect might arise from (at least) two sources, in isolation
or in combination: 
\begin{itemize}
\item cross-talk at the electronics level between VPol and HPol DAQ electronics traces, and/or
mutual inductance between the receiver antennas (the fact that the HPol antennas are in the null of the 
VPol beam pattern notwithstanding),
and/or
cross-polarized VPol response from a dominantly HPol receiver antenna, all of 
which should lead to observed H/V signals which are
simultaneous in time. This is, however, inconsistent with observations of local in-ice
VPol calibration pulser transmitter signals, at distances of approximately 30-50 meters from the 
receiver array, which show no evident HPol signals comparable to those observed here.
\item An ice-related effect, including:
\begin{itemize}
\item Inclined conducting layers within the ice, which act as an {\it in situ} polarizer.
\item ``circular'' birefringence, for which the birefringence basis is Left Circular Polarization vs. Right Circular Polarization, and which could ``rotate'' a pure VPol signal at the source to a mixture of VPol and HPol, resulting in, on average, an equal admixture of VPol and HPol propagating signals. The actual relative strengths of the two components would roughly vary linearly with distance. The fact that the HPol/VPol received signal amplitudes appear to be more-or-less constant, in all channels, for all receiver stations, disfavors this hypothesis.
\item ``linear'' birefringence, in which the signal projects onto two (presumably perpendicular) propagation axes, referred to as the
``ordinary'' (``O'', with a refractive index $n_O$ and a `fast' propagation velocity $c_0/n_O$, with $c_0$ the vacuum velocity of light) 
or ``extra-ordinary'' (``E'', with a 'slow' propagation velocity $c_0/n_E$) axes. Upon arrival at the receiver, each of the 
O and E signals then project back onto the receiver antenna axis, resulting in an expected doublet of signals for both V or H, with amplitude dependent on the inclination angle of
the underlying birefringent basis relative to ``true'' horizontal/vertical, and a separation time dependent on
the magnitude of the $n_O-n_E$ difference. 
\end{itemize}
\end{itemize}
We consider the last of these hypotheses in more detail in Section \ref{sect:biref}.

\message{Begin by considering the two Vpol antennas on string 2 for which there is only a vertical separation; we label these S2TV and S2BV for top and bottom Vpol. Each channel shows two pulses, one QD and one QR. The QD pulse on S2BV arrives before that on S2TV, indicating that the pulse is moving upward as it passes the station. Similarly, the arrival order for QR is reversed indicating a downward ray. These features support the interpretation illustrated by the ray tracing shown in Figure \ref{fig:rt}.}

\subsubsection{Comparison of observed D-R time difference against model}
The observed QD/QR time delays can be used to discriminate between putative n(z) models.
Several functional n(z) forms were tested against the measured timings. These included:
\begin{itemize}
\item Krav04: n(z)=1.37-(4.6$z$+13.72$z^2$) for $z>$-0.18 km, as suggested by a polynomial fit to direct radio wavespeed measurements at South Pole\citep{kravchenko2004situ}.
\item Model 2: n(z)=0.8+0.98/(1+exp(30z))
\item ARA: n(z)=1.78-0.43exp(13.2z),
\end{itemize}
with z in units of km, and increasingly negative
with increasing depth. The last two exponential forms match the density dependence expected in a gravitational field. 
For each putative model, we calculate the sum, over all 16 ARA channels, of the squared deviation between the measured
(QR,QD) arrival time difference, and the time difference predicted by a ray tracing model, assuming a given n(z) form, 
as shown in Table \ref{tab:fitsum}.
We find that the profile n(z)=1.78-0.43exp(13.2z)
currently used in the ARA Monte Carlo simulation provides the best fit to both the available density data (presented in \citep{barwick2018observation}), and also to the measured time differences between the QR and QD rays, observed in our experimental data. This functional
form also matches the $n_{surface}$(z=0)=1.35 and $n_{deep}$(z$<$-0.2)$\to$1.78 boundary conditions, consistent with 
density measurements at the South Pole. 

\begin{table}[htpb]
\begin{centering}
\begin{tabular}{ccc} \\ \hline
ARA station & Functional Form & $\Sigma(\delta_t^{QD-QR}(Model-Data))^2$ ($ns^2$) (IC22S/IC1S) \\ \hline
2 & Krav04 &  652230/1378220 \\ 
2 & Model 2 & 576483/1014030 \\ 
2  & ARA & 61752/23360 \\ \hline
3 & Krav04 & 1099209/487622 \\ 
3 & Model 2 & 974600/391978 \\ 
3 & ARA & 72309/43107 \\ \hline
\end{tabular}
\caption{Comparison of index-of-refraction model predictions to measured QD-QR time differences. Numbers in third column represent
sum, over all channels, of squares of differences between model-predicted (QD,QR) time difference relative to measurement, for IC22S and IC1S 
double pulses, respectively.}
\label{tab:fitsum}
\end{centering}
\end{table}

\begin{figure}[htpb]\centerline{\includegraphics[width=0.55\textwidth]{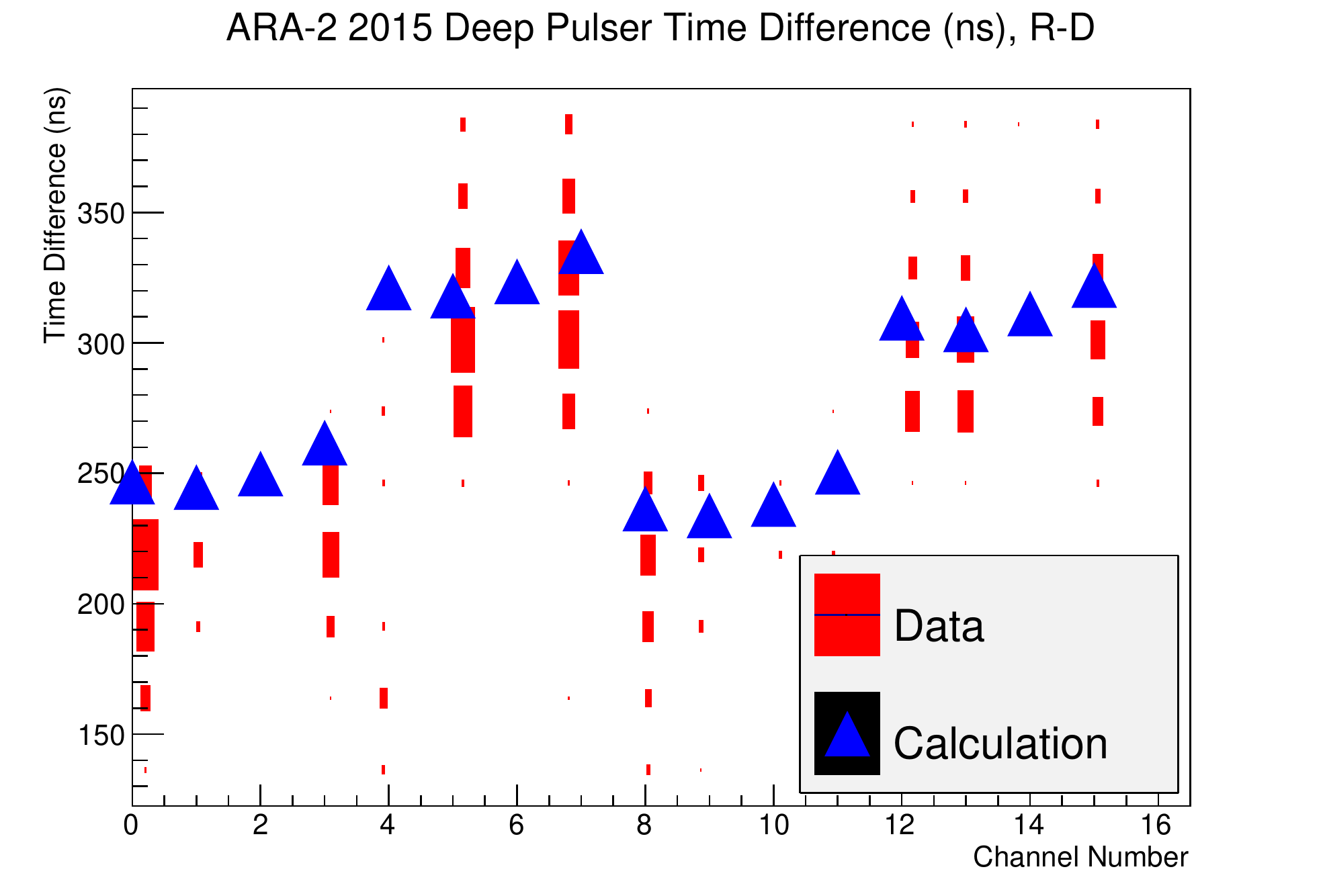}}\caption{Measured time differences between QD and QR rays for ARA-2 2017 deep pulser data, compared with calculation.}\label{fig:uzair}\end{figure} 
Channel-by-channel, Figure \ref{fig:uzair} compares the calculated, expected time difference between the arrival of the QD and QR rays
using our putative n(z) model (the default ARA model) with data.
Given the inherent uncertainty in discerning hit times algorithmically,
the two show acceptable agreement.

\subsection{QD vs. QR zenith angle direction reconstruction}
The pre-saturation leading-edges of the received waveforms are adequate to infer source timing, and therefore source location reconstruction. Using the calibrated station geometry, one can calculate the zenith angle of the arriving rays and verify their
consistency with the QR/QD arrival-angle hypothesis. For this reconstruction, we determine the incidence angles for the set of all QD rays, and the set of all QR rays, separately and 
independently.
\begin{figure}[htpb]\centerline{\includegraphics[width=0.55\textwidth]{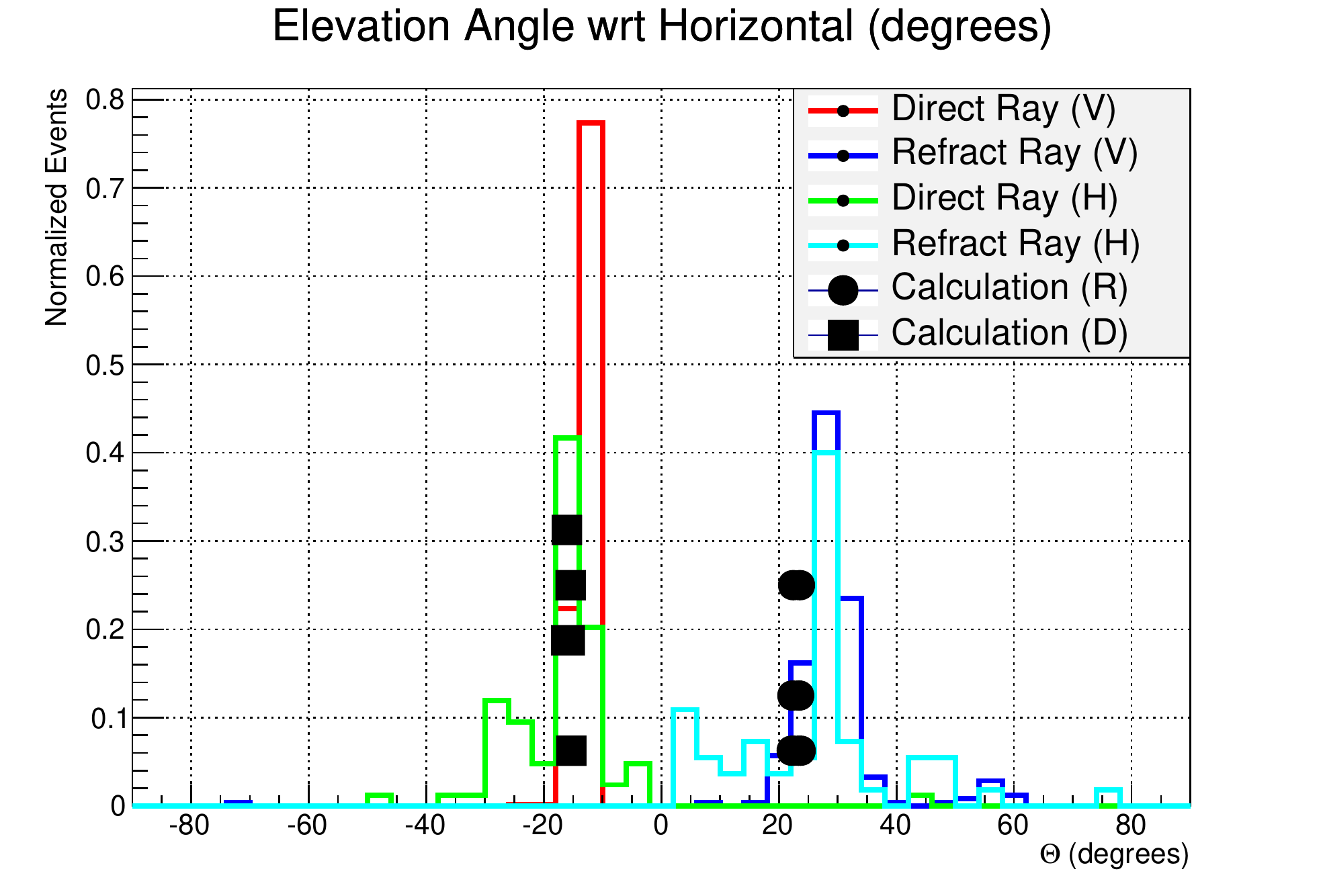}}\caption{Calculated zenith angle of incidence, relative to horizontal, for ARA-2 2017 deep pulser data. Histograms show the measured incidence angles for the QD and QR rays for both
VPol (red for QD and blue for QR) and also HPol (green for QD and cyan for QR) receiver channels. Points overlaid show expectation from model simulation, based on 
ARA index-of-refraction profile. We note an approximately 40 degree difference between reception angle for QD vs. QR.}\label{fig:AngleOfIncidence}\end{figure} 
As shown in Figure \ref{fig:AngleOfIncidence}, the earlier QD set of pulses arrive from below the horizontal; the later QR set of pulses arrive from above the horizontal, consistent with 
expectations for the QD and QR rays, respectively.

\subsection{Source Reconstruction in both Azimuth and Elevation}
\begin{figure*}[htpb]\centerline{\includegraphics[width=0.65\textwidth]{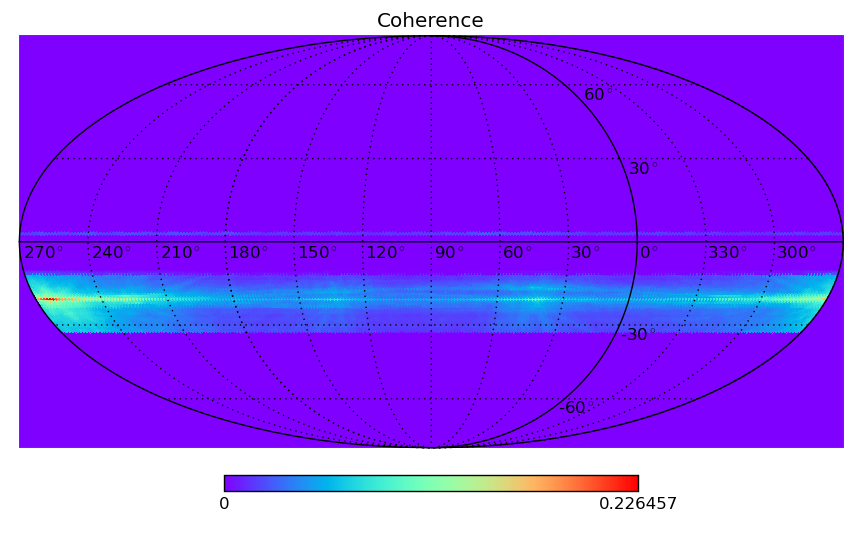}} 
\caption{Reconstructed interferometric image of ARA ICS22 deep pulser, using standard azimuth (horizontal) vs. zenith (i.e., elevation [vertical]) coordinates. Highest intensity pixel is visible in lower left.}\label{fig:IFG}\end{figure*}

\begin{figure}[htpb]
\includegraphics[width=0.45\textwidth]{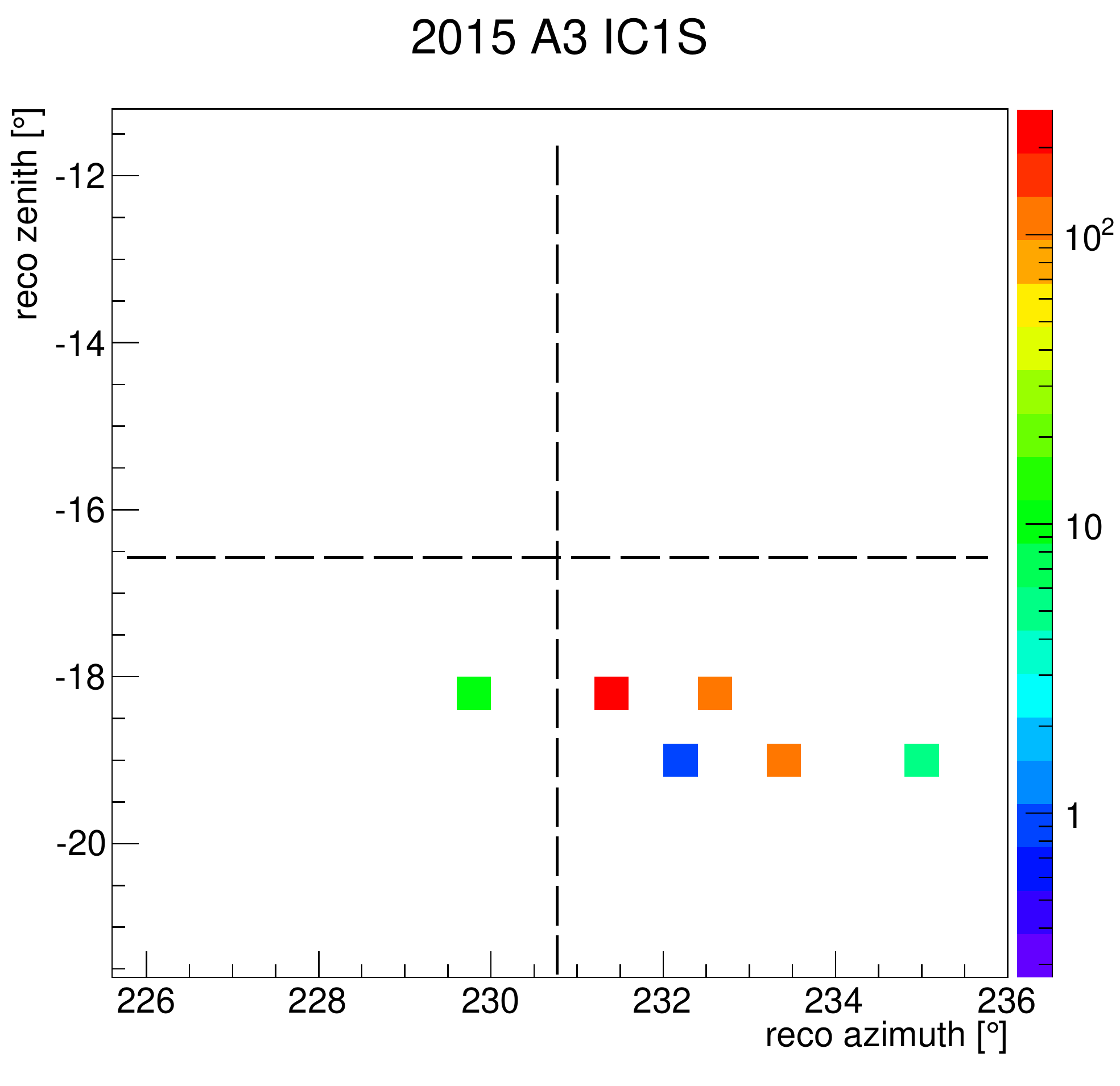}
\includegraphics[width=0.45\textwidth]{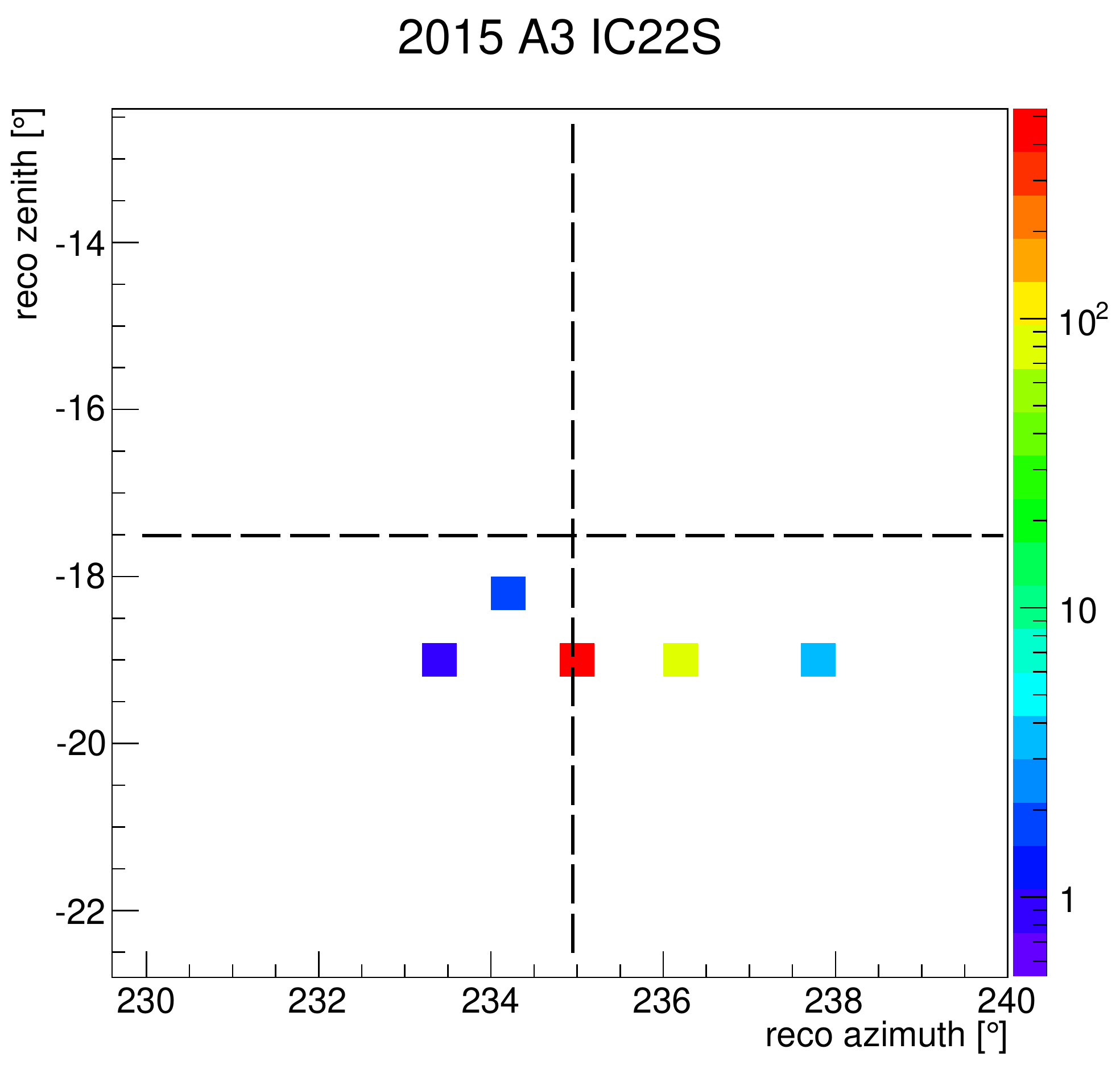}
\caption{ARA-3 azimuthal ($\phi$) vs. zenith ($\theta$) reconstructed source locations during time when ICS1 (left) and ICS22 (right) was pulsing. Dashed lines indicate ``true'' source location. There are two and zero outliers, respectively, not shown, beyond the plot boundaries.}\label{fig:ARA3_IC1}\end{figure}

\begin{figure}[htpb]
\includegraphics[width=0.45\textwidth]{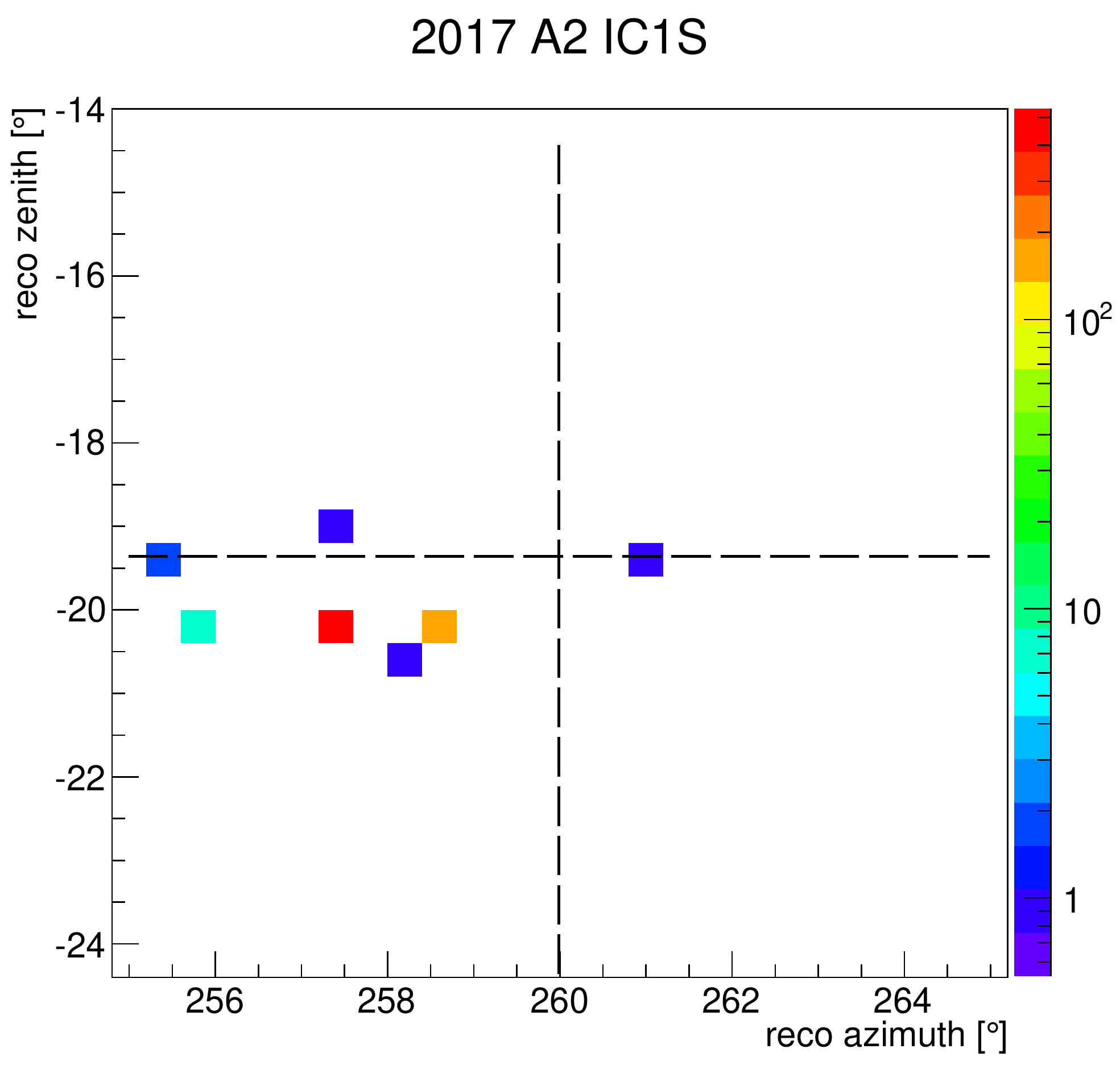}
\includegraphics[width=0.45\textwidth]{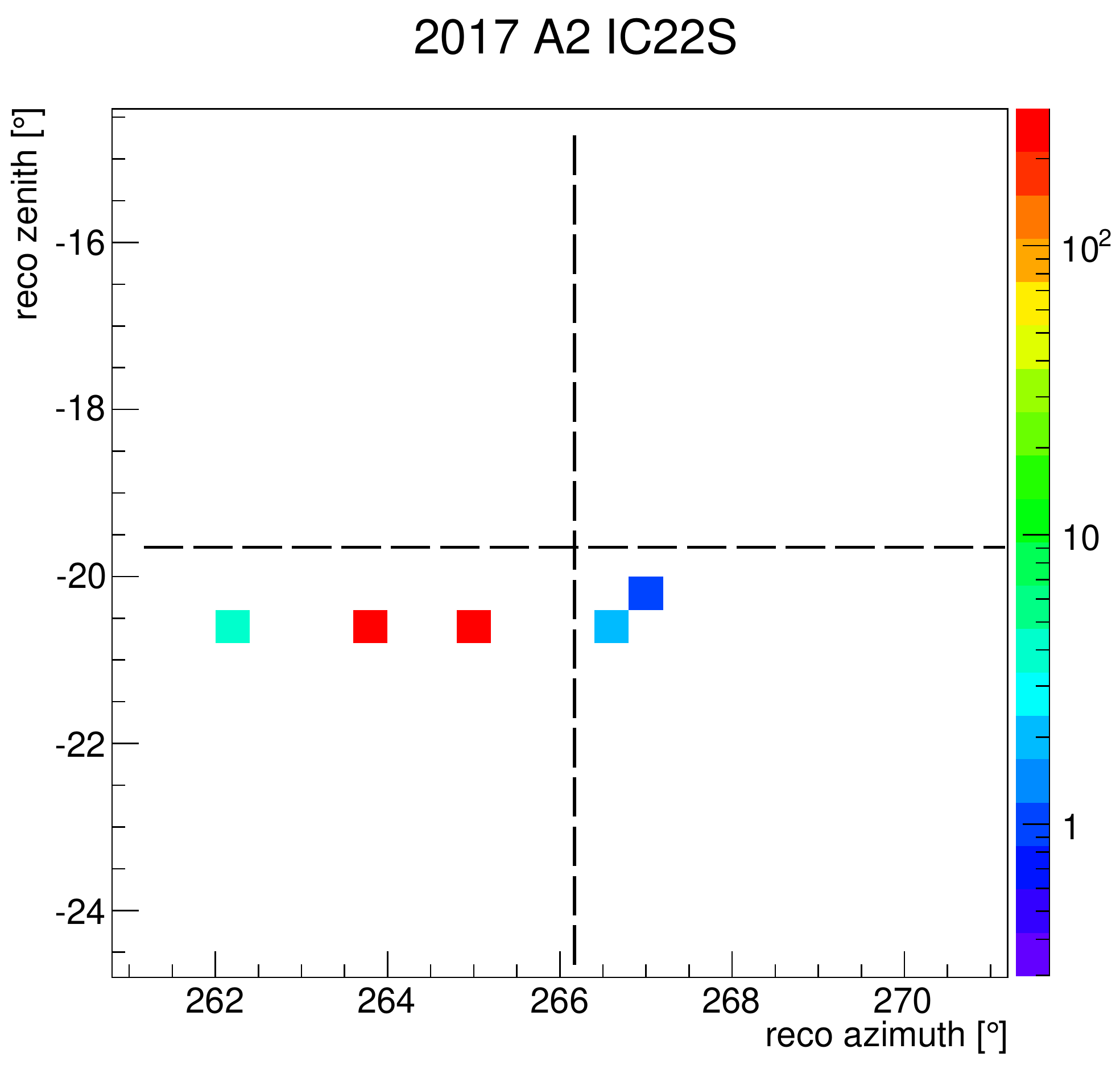}
\caption{ARA-2 azimuthal ($\phi$) vs. zenith ($\theta$) reconstructed source locations during time when IC1S (left) and ICS22 (right) were pulsing. 
There are 11 and 9 outliers, respectively, not shown, beyond the plot boundaries.}\label{fig:ARA2_IC22}\end{figure}

The standard ARA interferometry-based analysis event reconstruction\citep{allison2012design} cross-correlates the waveforms (QD$_{i}$, QD$_{j}$), where QD$_{i}$ and QD$_{j}$ are the quasi-direct signals on channels i and j. Through a fast lookup table of calculated, expected arrival times, a set of predicted $\delta$t$_{ij}$ (r, $\theta$, $\phi$) are produced for every possible source position. These are used to sample the Hilbert envelope of the i$-$j cross correlation function and give weights to the putative $\delta$t$_{ij}$\citep{behelerinterferometricICRC2017}. For this particular study, a second table for QR signal arrival times is built, and the cross correlation is extended to include all the available QR signals. Thus, the full reconstruction incorporates contributions from all possible QD$-$QD, QD$-$QR, and QR$-$QR pairs, and assesses the consistency of a putative 1 square degree source position in the sky with the ensemble of observed QD and QR hit times.
The source location map constructed for a deep pulser event is show in Figure \ref{fig:IFG}; the brightest $1^\circ\times 1^\circ$ pixel in that map is identified as the best-fit source position. The distribution of reconstructed deep pulser directions, as defined by the maximum intensity 
pixel in each event, is presented in Figures \ref{fig:ARA3_IC1} and \ref{fig:ARA2_IC22}.
In general, we reconstruct the source direction with accuracy 1--2 degrees in both azimuth and elevation.

\subsection{Range Reconstruction}
The reconstruction from QD rays permits a good direction to the vertex, but determining the
distance to the vertex is more difficult, as (absent additional QR information) this requires a determination of the curvature of the radiation
front, which is limited by both the modest 20 m baseline of the station and the inherent signal arrival time resolution.
There is $\sim$100 ps resolution in the time-difference between hit times, as determined by waveform cross-correlations. If the antenna locations, the index-of-refraction profile, individual channel cable delays and antenna group delays were all exactly known, the wavefront reconstruction would be entirely determined by this 100 ps timescale. Unfortunately, there are significant uncertainties in all of these quantities, which have a combined error at least an order of magnitude larger than the 100 ps timescale. 

In the case where both the QD and QR signals are observed, however,
comparison of the QD with the QR signal arrival times allows improved estimation of 
range-to-vertex without use of wavefront curvature information. Considering the quasi-reflected ray as if it were detected by
an ``image'' station above the ice surface, for a 200m deep receiver, the baseline for reconstruction by both rays is of order
twice the depth of the station, or 400m, enabling a full 3D reconstruction. 

\begin{figure}[htpb]
\includegraphics[width=0.45\textwidth]{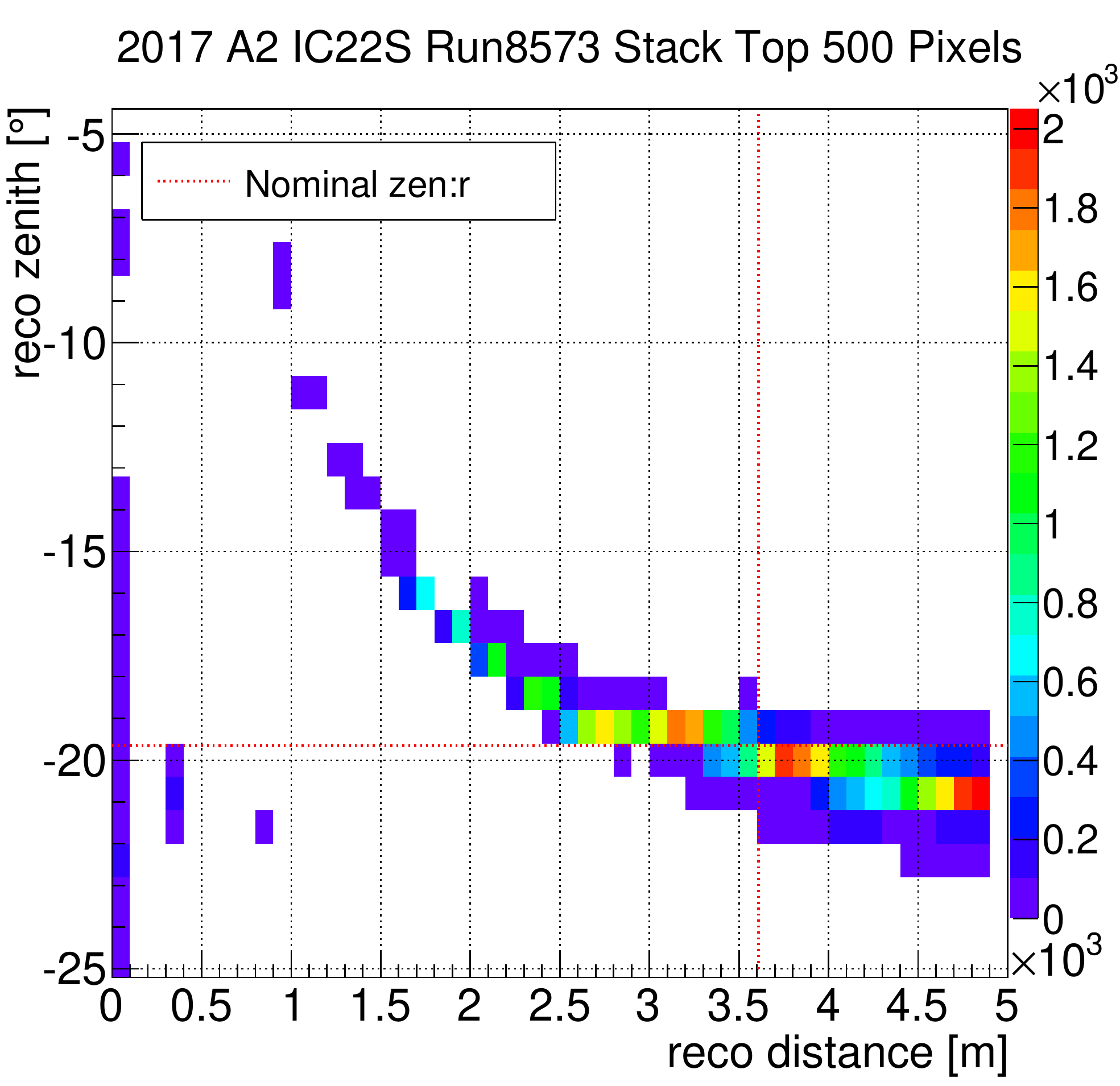}
\includegraphics[width=0.45\textwidth]{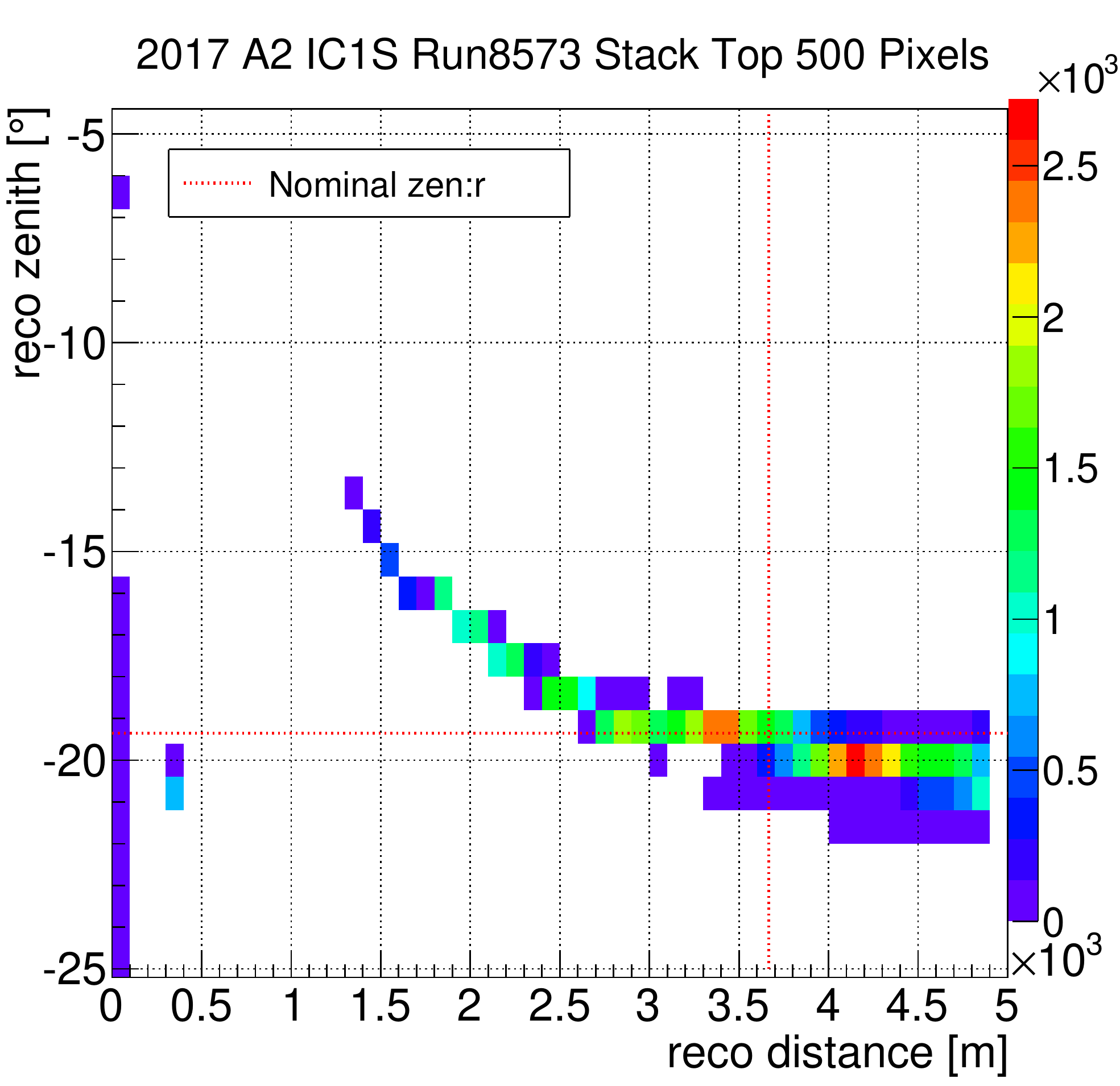}
\caption{ARA-2 ``stacked'' pixels range vs. elevation($\theta$) distribution. Ranking by source likelihood, we plot the (range, $\theta$) of the highest 500 pixels of each event. Shown is this stacked distribution from events during time when ICS22 was pulsing. The color intensity scale indicates the frequency of a putative (range, $\theta$) point.}
\label{fig:ARA2_range_IC22}\end{figure}

Similar to the procedure followed for interferometric reconstruction of azimuth and elevation, we can empirically determine the source range most consistent with the
observed QD-QR time difference.
The results of this exercise are presented in Figure \ref{fig:ARA2_range_IC22} for ARA-2 reconstruction of 
the ICS1 and
ICS22 pulsers. We observe that deviations in elevation reconstruction of order one degree typically translate into range errors of tens of percent.
Table \ref{tab:summary} summarizes our results numerically, and indicates that we can use the time difference between QD and QR signals to estimate the range to within $\sim$15\%.


\begin{table*}[ht]
\begin{center}
\begin{tabular}{c|ccc} \\ 
 & \multicolumn{3}{c}{{\tt Nominal}/{\bf Reconstructed}} \\ \hline
 Station/Source & $\theta$ (deg) & $\phi$ (deg) & r (m) (deviation) \\
A2/IC1S & {\tt -19.36}/{\bf -20.19$\pm$0.07} & {\tt 259.99}/{\bf 257.6$\pm$0.56} &  {\tt 3666$\pm$543}/{\bf 4215} (+15\%) \\ 
A2/IC22S & {\tt -19.65}/{\bf -20.64$\pm$0.02} & {\tt 266.17}/{\bf 264.4$\pm$0.64} & {\tt 3609$\pm$530}/{\bf 4896} (+36\%) \\ 
A3/IC1S & {\tt -16.57}/{\bf -18.21$\pm$0.35} & {\tt 230.77}/{\bf 232.2$\pm$0.95} & {\tt 4269$\pm$696}/{\bf 4711} (+10\%) \\
A3/IC22S & {\tt -17.51}/{\bf -18.9$\pm$0.05} & {\tt 234.95}/{\bf 235.2$\pm$0.45} & {\tt 4040$\pm$603}/{\bf 4298} (+6\%) \\
\hline
\end{tabular}
\caption{Summary of reconstructed deep pulser source locations compared with ``known'' source location. For
each pair presented, the first value represents the nominal (`true') value; the second value (in bold) represents the value our source reconstruction algorithm returns, along with the associated statistical error.}
\label{tab:summary}
\end{center}
\end{table*}

\section{\label{sect:biref}Study of H/V relative signal timing and evidence for birefringence}
\begin{figure}[htpb]\centerline{\includegraphics[width=0.55\textwidth]{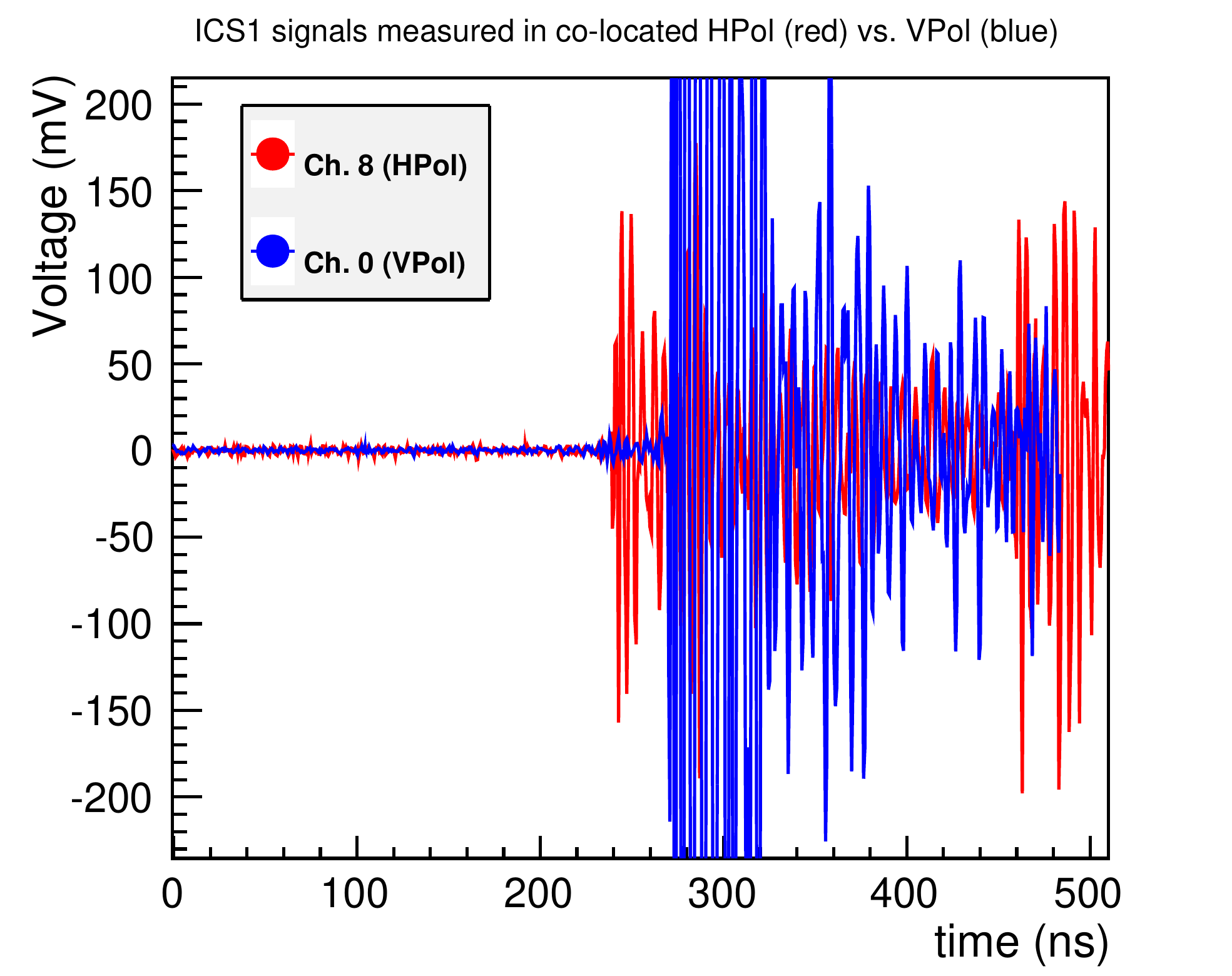}}\caption{Zoom of channels 0 (VPol) and 8 (co-located HPol) 
for signals received from IC1S, illustrating time difference between HPol and VPol waveforms.}\label{fig:ARA2_Ch0_Ch8}\end{figure}

The pathlengths from the ICS1 and ICS22 transmitters to the ARA receiver stations is of order 3--5 km, along a predominantly horizontal trajectory. To our knowledge, this is
the longest horizontal baseline used for testing electromagnetic signal propagation, of any type, in the polar regions. As such, this provides a singular opportunity to
probe the wavespeed variation with polarization. As shown in Figures \ref{fig:ARA2_DP4} and \ref{fig:ARA3_DP5}, and zoomed for two ARA-2 channels in Figure
\ref{fig:ARA2_Ch0_Ch8}, the HPol signal from the deep pulsers is clearly advanced by
20--30 ns, for both ARA-2 and ARA-3, relative to the signal registered in the co-located VPol receiver. Since the elevation angles of the signal arrival are within $\sim$20 degrees of horizontal (Figure \ref{fig:AngleOfIncidence}),
the expected signal arrival times should be nearly identical for this H/V channel pair.

Given that ice crystals are known to exhibit linear, but not circular birefringence\citep{hargreaves1978radio},
we consider the arrival of the HPol signals prior to the arrival of the VPol signals to be most plausibly
explained by linear ice birefringence, although this requires some bulk ice crystal alignment.
In the absence of any ice crystal directional asymmetry, one would expect the radio-frequency wavespeed to be 
uniform in all directions. Isotropy of the Antarctic ice sheet is broken in two directions -- vertically, due to 
the gradient in hydrostatic pressure with depth and resulting in a compressional stress on ice crystals,
and horizontally, due to 
the local ice flow direction and resulting in a torsional strain, and therefore, a preferred axis laterally. 
\message{For example, for both ARA-2 and ARA-3, the VPol antenna is about 3m below the Hpol antenna, and yet the upcoming QD signal arrives at the Hpol antenna about 30 ns ahead of the Vpol signal. Similarly, the QR pulse is also seen on Hpol antennas and also leads. In addition, the Hpol channels show a secondary pulse which appears at about the same time as the corresponding Vpol.} 
\message{The peaks of the H/V envelope function are displaced by about 30 ns from the expected arrival time separation. In addition to the displaced main peak of the envelope, both correlations show a secondary peak near the expected arrival time, a feature consistent with the direct observation of the waveforms. The situation is somewhat obscured, however, by features of the main Vpol pulse. The deep pulser waveforms have a width of about 42 ns, with pronounced leading and trailing edges which create features in the correlation functions. These features can obscure possible features from a second pulse. The result is an H-V envelope function that has two wings at $\pm$30--40 ns, but with asymmetric amplitudes.}

\begin{figure*}[htpb]
\centerline{\includegraphics[width=0.8\textwidth]{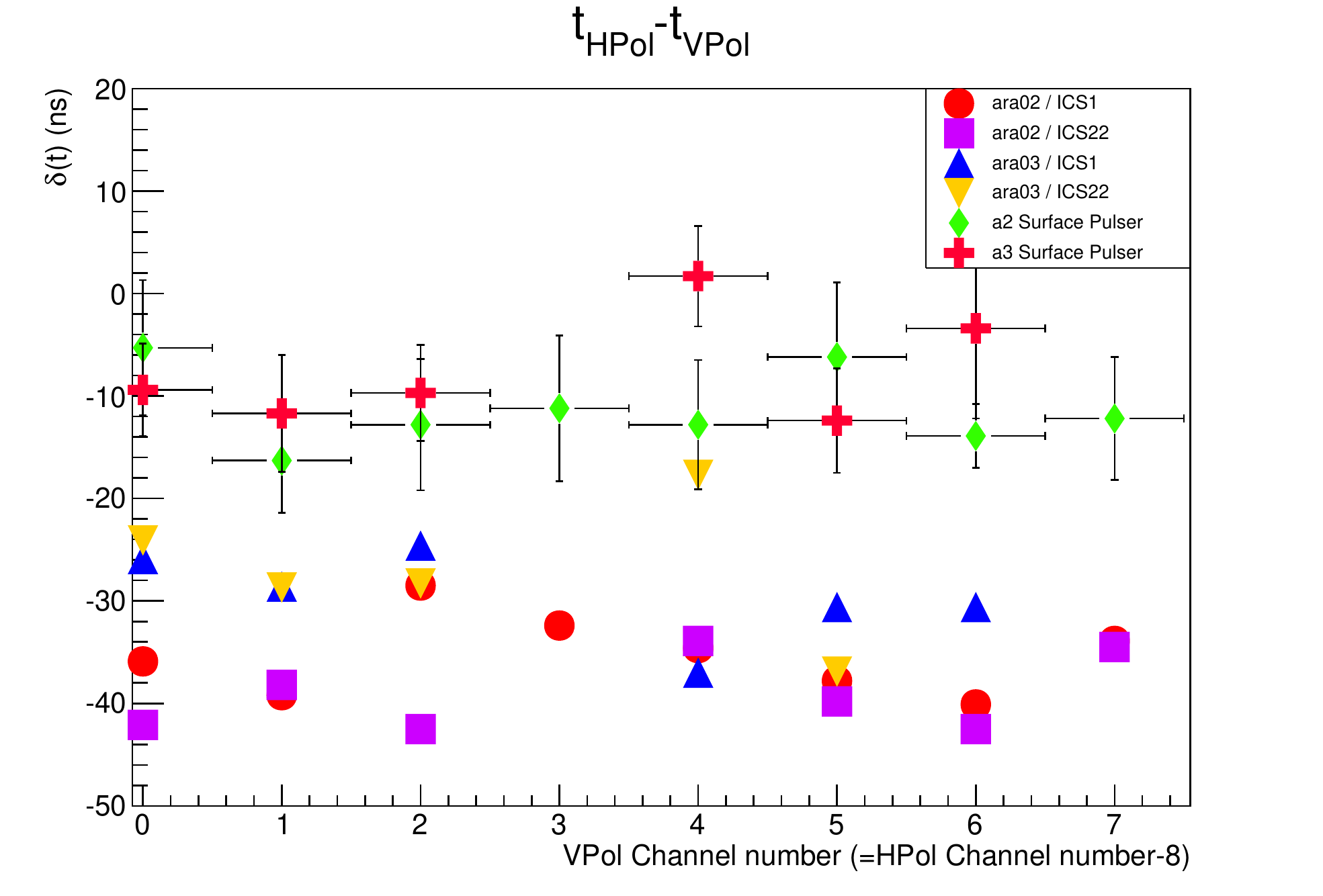}} 
\caption{$\delta_t^{V-H}$ ARA-2 and ARA-3 data. x-axis refers to channel number
of vertically-polarized receiver; y-value gives the (corrected) time difference between registered signal arrival time on 
the co-located HPol channel for that hole/station relative to registered VPol signal arrival time for an H/V pair. For comparison,
results from a surface transmitter survey, at lateral distances 100--200 meters from each station, are also shown.
In the absence of birefringence and/or timing miscalibration, all data points should approximately lie on the x-axis, consistent with zero 
propagation-time difference. We observe that
ARA-2 has typical time differences of 30-40 ns relative to a propagation time of approximately 20 $\mu$s; for ARA-3, values are approximately 5-10 ns smaller. Systematic errors on ARA2 and ARA3 H/V time differences are indicated by surface pulser survey data, and
estimated at 3--4 ns.}
\label{fig:sum}
\end{figure*}

To calculate the magnitude of the birefringent asymmetry between H/V pairs in data, we must correct for the 
2-3 meter shallower deployment depth of the H-pol antenna of the pair, leading to a timing 
correction $\delta_t^{corr}\approx n(z)\delta z(sin\theta_i)/c_0$, with $\delta z$ the vertical separation
of the HPol vs. VPol receivers, $\theta_i$ the ray incidence angle relative to the horizontal, $n(z)$ the local
index-of-refraction, and $c_0$ the velocity of light in vacuum. After applying this correction,
the calculated HPol advance relative to same-string VPol channels is summarized in Figure \ref{fig:sum}, for deep pulser
signals observed in ARA-2 and ARA-3. 
In the Figure, the horizontal value corresponds to the $i^{th}$ VPol receiver channel; the y-value
gives the measured time difference, after correction, relative to the zero birefringence expectation, for the
nearest HPol channel.
We note that prior to the vertical-displacement correction described above, the ARA-2 and ARA-3 $\delta_t^{V-H}$ values are consistent with each other;
after correction, the value of $\delta_t^{V-H}$(ARA-3) is typically 15--20\% smaller than for ARA-2, despite the fact that
the propagation distance from the deep pulsers to the ARA-3 station is approximately 15\% longer. 
A simple explanation for this is, of course, that the H-V pair vertical separations have been mis-tabulated by $\sim$1.5 meters.
Alternately, in the birefringent model, this result may suggest 
that the vector from the deep pulsers to ARA-2 is more aligned with the underlying birefringent basis than ARA-3. 
Geometrically, the line from IC1 to ARA-2 is very nearly directly perpendicular to the horizontal ice flow
direction; the line from IC1 to ARA-3 is offset by $\sim$30 degrees relative to that line (Figure \ref{fig:a0-a3}). 

\subsection{Cross-checks}
Aside from birefringence, the observed time delay between the arrival of the HPol vs. VPol signals could, in principle, be due to a simple mis-calibration of the HPol vs. VPol receivers. To
test this, we considered data taken during an RF survey of stations 2 and 3 for which a surface transmitter was broadcast from a dipole, with its long axis inclined at an angle of 45 degrees relative to vertical, within 200 m laterally of each station.
Given the transmitter inclination angle, these data should have roughly equal amounts of HPol and VPol broadcast signal. We observe 
in Figure \ref{fig:sum} an offset between H and V of approximiately
--5 -- --10 ns, consistent in sign, but of somewhat smaller magnitude than we observe for broadcasts from the deep pulsers.\footnote{This effect is, in principle, consistent with horizontal birefringence through the upper firn layer of the ice sheet.} 
From inspection of the waveforms themselves, we observe a sharper rise time for the VPol signals, which we attribute to 
the smaller bandwidth of the HPol antennas compared to VPol antennas; however, this effect should have the result of staggering, rather than
advancing the peaks of the HPol signals relative to VPol. Therefore, 
pending additional investigation, we interpret these data as 
conditional evidence for birefringence for signals propagating from the $\sim$1450-m deep pulsers.

The evidence for birefringence must be interpreted in the context of previous results. 
The RICE Collaboration observed birefringence in vertical propagation of rays which reflect 
off the bedrock\citep{Besson:2010ww}. They conclude that the time delays accumulate mostly in the deep ice, below about 1200 m, as they observe
no evidence for birefringence in the upper ice.
This is consistent with a crystal orientation fabric
(COF) determined by shear in the ice flow\citep{price2002temperature}, which is rather modest in the upper ice. The current
results, however, suggest that the upper ice does exhibit birefringence for horizontal propagation. 
If the COF of the upper ice
is dominated by gravity, then a vertical ray would not exhibit birefringence, but a horizontal ray
would since the Vpol would be along the net $c$-axis while Hpol would be transverse to it.

\section{Conclusions and Discussion}
\message{At the same time, close examination of Figure 3 shows that while the QD rays have sharp leading edges, the QR rays have precursors consistent with a picture where the path for the upper part of the QR ray can be shortened via a horizontal propagation of part of the signal power.}
As shown herein, the time difference between the registration of direct vs. refracted (or reflected) rays permits an
estimate of the range-to-vertex, once the azimuth and elevation of an incident signal has been determined
through interferometry. Additionally, if the time delay between signal arrival times for
HPol vs. VPol receivers ($\delta_t^{V-H}$) can be quantified for all geometries, then an additional 
constraint on the event geometry may be afforded by the measured magnitude of birefringent time difference.

At South Pole, the ARA strategy thus far has been based on receiver deployment at approximately 200 meters.
In evaluating the optimal depth to site a future radio receiver array, one must consider the following:
\begin{itemize}\item The launch angle difference between the QD and the QR rays increases 
as the receiver depth $z_{Rx}$  also increases. The data presented here are based on reception 
of signals from an in-ice transmitter dipole with a very broad beam pattern, however, the Cherenkov cone from a neutrino has only a 1--2 degree transverse thickness.
\item As the receiver depth $z_{Rx}$ increases to 200 meters, the measured time difference between reception of the QD and the QR rays also increases, approaching one microsecond, and therefore requiring $\sim$2$\mu$s of data buffering, per channel, to ensure capturing both the QD and the QR rays.
\end{itemize}
These considerations, combined with the logistical overhead associated with drilling ice-holes down to 200 meters, suggest that a shallow receiver array, 
deployed at a depth of 25--50 meters may retain high efficiency for observation of the QD and QR rays from a neutrino interaction, have good (of order hundreds of meters) range resolution, and sufficient timing resolution to measure $\delta_t^{V-H}$ (giving a redundant measure of range-to-vertex) within a tight (512 ns) waveform capture time window.

The observations presented here are a small part of the radio-glaciological data needed to fully
characterize RF propagation of relevance to the ARA experiment. In this regard, within the last two
years, a 1700-m deep ice core, 0.5--several km from current ARA stations, 
was extracted from the South Pole ice\citep{casey20141500}. 
With the consent of the NSF, a piezo-electric technology radio-frequency transmitter, based on the same model as that employed in the
balloon-borne HiCal experiment\citep{gorham2017antarctic}, was used in January, 2018 to
sample the upper 800 meters of the SPICE borehole. The beam pattern of that transmitter is somewhat more isotropic than that used in 
the deep pulser studies described herein, and, by design, contains significant HPol as well as VPol content. Those data are currently under analysis.
In parallel, we are working to develop a first-principles model, based on measurements of the crystal orientation fabric, that absolutely predicts, for
any polarization and $k$-vector, an observed birefringent asymmetry.


\section{Acknowledgments} 
We thank the National Science Foundation for their generous support through Grant NSF OPP-1002483 and Grant NSF OPP-1359535. We further thank the Taiwan National Science Councils Vanguard Program: NSC 102-2628-M-002-010 and the Belgian F.R.S.-FNRS Grant4.4508.01. We are grateful to the U.S. National Science Foundation-Office of Polar Programs and the U.S. National Science Foundation-Physics Division. We also thank the University of Wisconsin Alumni Research Foundation, the University of Maryland and the Ohio State University for their support. Furthermore, we are grateful to the Raytheon Polar Services Corporation and the Antarctic Support Contractor, for field support. A. Connolly thanks the National Science Foundation for their support through CAREER award 1255557, and also the Ohio Supercomputer Center. K. Hoffman likewise thanks the National Science Foundation for their support through CAREER award 0847658. A. Connolly, H. Landsman, and D. Besson thank the United States-Israel Binational Science Foundation for their support through Grant 2012077. A. Connolly, A. Karle, and J. Kelley thank the National Science Foundation for the support through BIGDATA Grant 1250720. B. A. Clark thanks the National Science Foundation for support through the Graduate Research Fellowship Program Award DGE-1343012. D. Besson and A. Novikov acknowledge support from National Research Nuclear University MEPhi (Moscow Engineering Physics Institute). R. Nichol thanks the Leverhulme Trust for their support.
\bibliography{aZref}
\bibliographystyle{unsrt} 

\end{document}